\documentclass[12pt,english]{article}
\usepackage[T1]{fontenc}
\usepackage[latin9]{inputenc}
\usepackage{geometry}
\usepackage{float}
\usepackage{booktabs}
\usepackage{graphicx}

\makeatletter

\providecommand{\tabularnewline}{\\}

\newcommand{\lyxaddress}[1]{
	\par {\raggedright #1
	\vspace{1.4em}
	\noindent\par}
}

\date{}
\usepackage{gensymb}

\makeatother

\usepackage{babel}
\begin{document}
\title{Sintering BFO targets for sputtering}
\author{G. Orr\thanks{gilad.orr@ariel.ac.il}, A. Goryachev and G. Golan}
\maketitle

\lyxaddress{\begin{center}
Ariel University, Ariel, 40700 Israel
\par\end{center}}
\begin{abstract}
Ceramic $BiFeO_{3}$ samples were prepared by rapid sintering at $880\degree C$.
Two compositions were examined. A 56/44 $Bi_{2}O_{3}/Fe_{2}O_{3}$
mole\% composition and a $56Bi_{2}O_{3}\cdot44Fe_{2}O_{3}+6.5wt\%\,NaCl$
composition. The samples were heat treated at different times up to
8 minutes and the phase content was examined as a function of the
time using XRD measurements and analysis. It was demonstrated that
using both compositions, maximum $BiFeO_{3}$ phase content is obtained
after 3.5 minutes. In the former approximately 50\% of the material
transformed to $BiFeO_{3}$ while in the latter 98.5\%.
\end{abstract}

\section{Introduction}

BFO has regained much interest in recent years due to its multiferoic
nature \cite{spaldin2010multiferroics}. It is one of the few known
materials which simultaneously posses both ferroelectric and ferromagnetic
ordering at room temperature. It has promising applications in photovoltaics,
and due to the magneto-electric coupling between the electric and
magnetic polarization providing for new device design. As such, the
material shows much promise in realizing spintronic devices, sensors
and multistate memory devices \cite{spaldin2010multiferroics,catalan2009physics}.
It has been demonstrated \cite{lu2011phase} that impurities will
result in additional phases, the more common and stable ones being
the $Bi_{2}Fe_{4}O_{9}$ (mullite) and $Bi_{25}FeO_{40}$ (sillenite)
with a relatively short lived phase of $BiFeO_{3}$ (BFO). Of the
two variants ($BiFeO_{3},$$Bi_{2}Fe_{4}O_{9}$) much work has been
done in sintering ceramics composed of the variety of phases with
the aim of obtaining materials which are composed mostly of $BiFeO_{3}$
nano crystallites. This is required for fabricating experimental devices
based on deposition of thin films such as spintronics based components
and sensors. $BiFeO_{3}$ sintered ceramic disks may be used as targets
for the thin film deposition process.

While all three phases appear at different ratios within the sintered
material we can expect that those phases with the greater change in
Gibbs energy will be more stable. Based on experimental work, Phapale
et al. \cite{phapale2008standard} evaluated the heat capacity and
derived the standard Gibbs energy of formation for the above compounds
from room temperature up to $640\degree C$. Selbach et. al. expanded
this work demonstrating that at the temperature range of $447\sim767\degree C$
Gibbs energy of formation of $BiFeO_{3}$ is a metastable compound
which would be the first to nucleate but would eventually transform
into $Bi_{2}Fe_{4}O_{9}$ and $Bi_{25}FeO_{40}$. But above $767\degree C$
increasing to the proximity of the peritectic temperature at $930\degree C$,
or below it is a stable compound. Carvalho et. al. \cite{carvalho2008synthesis}
using sol-gel combustion to create ceramic samples of $BiFeO_{3}$
have also demonstrated similar results to Selbach's work. They further
demonstrate that at a temperature of $600\degree C$ the $BiFeO_{3}$
transforms into the more stable mullite phase $Bi_{2}Fe_{4}O_{9}$
over many hours. This is the reasoning behind the short period heating
and rapid cooling techniques for obtaining BFO ceramics \cite{mukherjee1971kinetics,wang2004room,awan2011synthesis}.
Using ultra pure starting materials Lu et. al. \cite{lu2011phase}
tested the stability as a function of the crucible type (gold or aluminum
oxide) and the process from which the compound was synthesized. The
processes consisted of solid state reaction of the starting materials
without reducing parasitic phases using $HNO_{3}$, solid state reaction
of the materials followed by parasitic phase reduction and crushed
crystals. Composition stability was tested at $850\degree C$ for
24 and 48 hours and $855\degree C$ for 24 hours. Both temperatures
are above $767\degree C$ so we should expect them to form the more
stable $BiFeO_{3}$ phase. After 48 hours at $850\degree C$ both
the phase reduced (alumina crucible) and the crushed crystal (gold
crucible) did not decompose, while after 24 hours 44\% of the non
reduced compound decomposed. At $855\degree C$ after 24 hours, 7\%
of the crushed crystals in gold crucible, 26\% of the parasitic phase
reduced in gold crucible, and 87\% of the parasitic phase reduced
in alumina crucible decomposed. Evidently, and as expected at $850\degree C$
BFO is the stable phase, at the same temperature the material without
parasitic phase reduction decomposed considerably. The important point
to consider is that the $HNO_{3}$ treatment dissolves every compound
but the BFO and the $Bi_{2}Fe_{4}O_{9}$, thus without such a treatment,
some phase impurities and not necessarily only the sillenite phase
exist resulting in an increased decomposition. At $855\degree C$
the volatility of bismuth comes into play. Lu claims it is a phase
transformation but we doubt it as no one has observed a phase transformation
at that temperature, and that includes the DSC results that they present
in the mentioned article. It is more likely a compositional change
due to bismuth evaporation. Considering the parasitic reduced phase
in the alumina crucible, something totally different is happening.
According to the published phase diagrams \cite{speranskaya1965phase,maitre2004experimental,palai2008beta,lu2011phase}
we should not expect the mullite phase unless we exceed $930\degree C$
or if we suffer from considerable evaporation of bismuth. Hence, we
are drawn to suspect that the existence of the mullite phase in areas
which it is not a favorable phase is indicative of impurities. In
order to test this assumption we will compare a $Bi_{2}O_{3}$ saturated
composition against the same $Bi_{2}O_{3}$ saturated composition
but with $NaCl$ as an impurity.

\section{Experimental}

\subsection{Material synthesis and sintering}

The starting materials for the samples which were sintered are based
on the what was found to be the optimal composition by Bush et. al
and Gabbasova et. al \cite{gabbasova2010bi} for growing BFO macroscopic
crystals from the melt. The melt composition is 75.6 weight\% $Bi_{2}O_{3}$,
17.9 weight\% $Fe_{2}O_{3}$, and 6.5 weight\% $NaCl$. Disregarding
the NaCl that provides the sodium as a spectator ion, this translates
to 56 mole\% $Bi_{2}O_{3}$ and 44 mole\% $Fe_{2}O_{3}$. We regard
the above material as the material with controlled impurities. This
was compared to results obtained with a composition of 78.85 weight\%
$Bi_{2}O_{3}$and 21.15 weight\% $Fe_{2}O_{3}$, considered the material
without controlled impurities. It gives ample room for study without
the system shifting into a different phase system during the study
due to mass loss. A recent article \cite{sharma2019impact} illustrates
this mass loss using DTA/TGA curves stating that above $400\degree C$
and up to $800\degree C$ approximately 2.5\% Bi evaporation occurs.
This corresponds to approximately a 1 mole\% decrease in Bi content.
Based on \cite{lu2011phase} we can assume that at $855\degree C$
it increases to 7 mole\% over a period of 24 hours. As shall be seen,
the above compositions will assist us in evaluating dynamics and evolution
of the different compounds in the sintered samples. The constituents
were mixed and milled, followed by calcination at $800\degree C$
for 4 hours in alumina crucibles. The furnace was heated at a rate
of 100 degrees per hour with the material inside until it reached
the target temperature. After 4 hours of calcination it was cooled
to room temperature at a rate of 100 degrees per hour. The calcinated
material was ground for 2 hours using a ball grinder and sieved through
a $50\mu m$ mesh. Approximately $0.6g$ was pressed uniaxially into
$12.7$ mm disks with a pressure of approximately 100MPa. Pressed
disks were inserted into a muffle furnace set at $880\degree C$ for
a certain period of time set in minutes and were extracted followed
by quenching in air. 

\subsection{Evaluation of the calcined materials}

\begin{figure}[H]
\centering{}\includegraphics[scale=0.3]{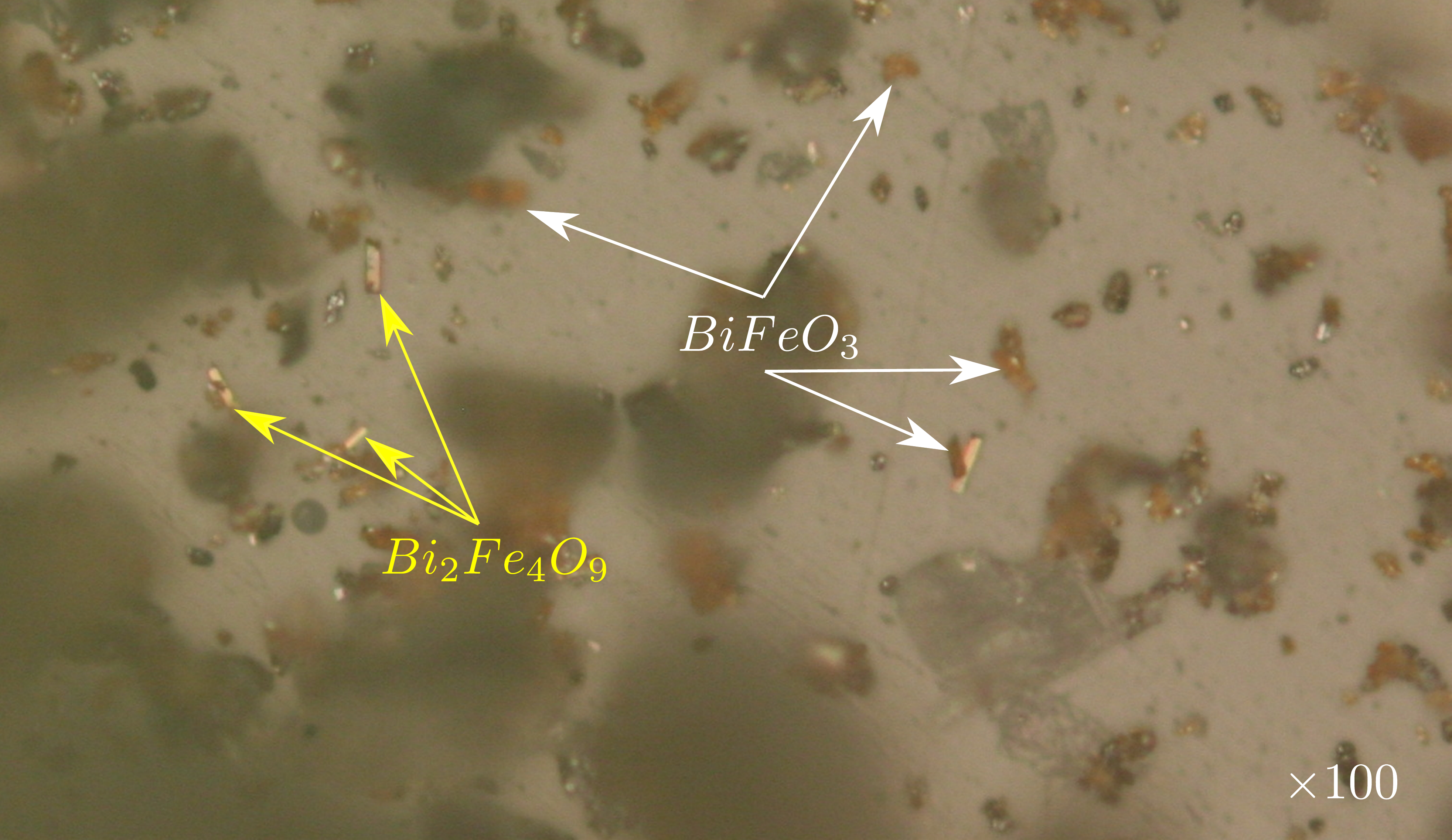}\caption{\label{fig:Fragments-of-crushed}Fragments of crushed material after
calcination prior to grinding. Much of the material has crystallized
into its distinguishable phases.}
\end{figure}

\begin{figure}[H]
\begin{centering}
\includegraphics{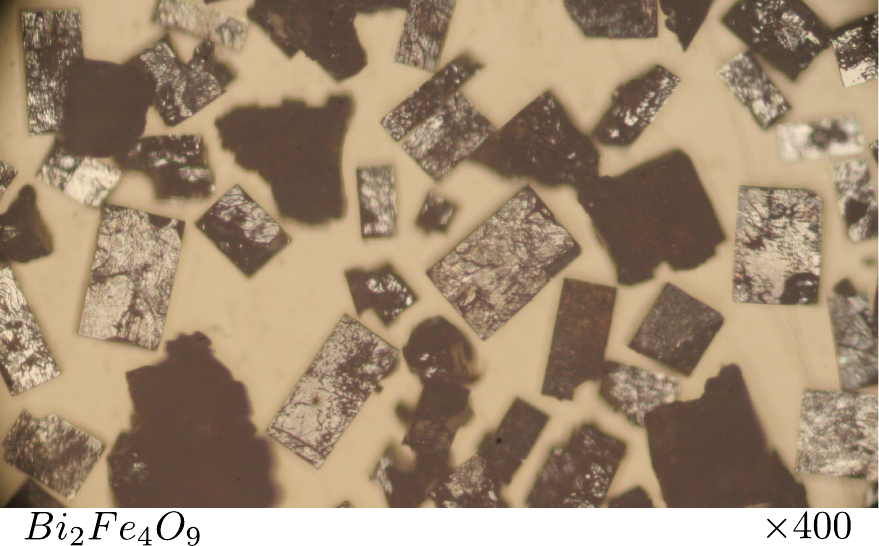}
\par\end{centering}
\begin{centering}
\includegraphics[scale=0.5]{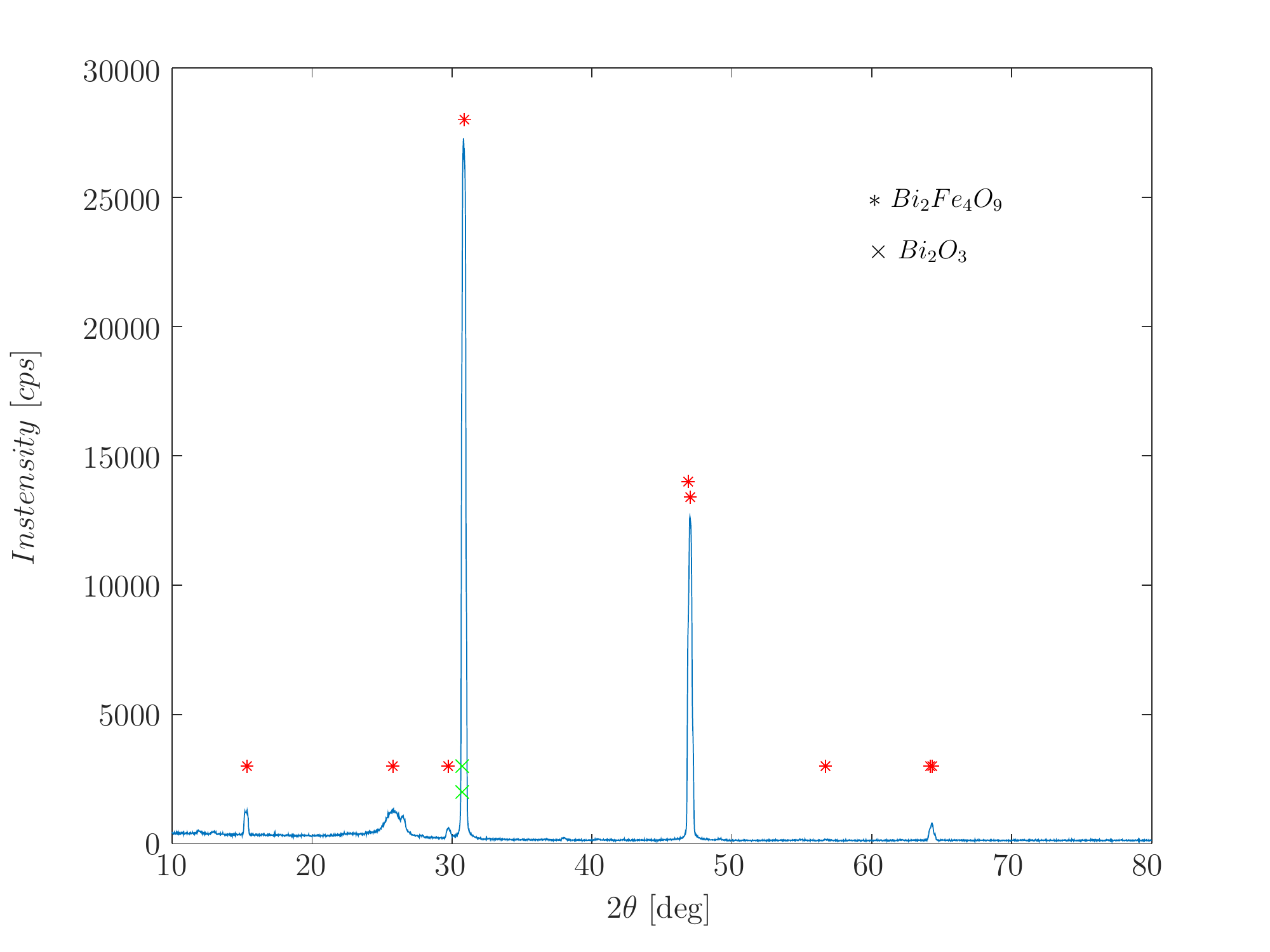}
\par\end{centering}
\caption{\label{fig:-crystals-extracted}$Bi_{2}Fe_{4}O_{9}$ crystals extracted
from the calcinated material with its XRD profile}
\end{figure}

The resulting calcined materials and sintered samples were recorded
visually using a metallurgical microscope, followed by XRD and SEM
analysis. The XRD measurements were conducted on a Rigaku SmartLab
using the SmartLab Studio II software for analysis. One of the most
important aspects for comprehending the optical microscope images
is the ability to distinguish the morphology of at least one of the
phases. As we shall see from the XRD analysis, the three obtained
crystalline phases important to us, are $BiFeO_{3}$, $Bi_{2}Fe_{4}O_{9}$
and $Bi_{25}FeO_{40}$. Based on the Gibbs energy of formation \cite{selbach2009thermodynamic}
during the initial stages of nucleation, and for a relatively short
time a metastable $BiFeO_{3}$ compound prevails over the $Bi_{2}Fe_{4}O_{9}$
and the $Bi_{25}FeO_{40}$. It means as well, that the metastable
compound will prevail at a lower temperature range. In the longer
time span we expect the more stable compounds as are described by
the $Bi_{2}O_{3}-Fe_{2}O_{9}$ phase system. For the 56/44 mole\%
Bi/Fe composition we expect either BFO or sillenite. We begin by analyzing
the compounds after calcination. Figure \ref{fig:Fragments-of-crushed}
illustrates an x100 optical microscope image of fragments of crushed
calcinated material before grinding.
The figure illustrates that during the calcination process some of
the constituents may be separated for analysis. In the example illustrated
in the figure two phases are distinguishable, the metalic colored
orthorhombic $Bi_{2}Fe_{4}O_{9}$ and the orange tinted brown rhombohedral
$BiFeO_{3}$. Crystal phases developed due to the calcination process
were analyzed by crushing some of the calcinated compounds rinsing
them in $HNO_{3}$ and washing them in water. This procedure dissolved
all the different phases but the $Bi_{2}Fe_{4}O_{9}$ and $BiFeO_{3}$.
The two remaining crystals were manually separated and were analyzed
with XRD. Figure \ref{fig:-crystals-extracted} illustrates an example
of separated $Bi_{2}Fe_{4}O_{9}$ crystals.
For the above example using split pseudo-Voigt fitting of the diffraction
peaks and fitting the peaks with the ICDD database. The lattice parameters
were measured to be $a=7.9947\mathring{A};\,b=8.4599\mathring{A};\,c=5.9254\mathring{A}$. 

\begin{figure}[H]
\begin{centering}
\includegraphics{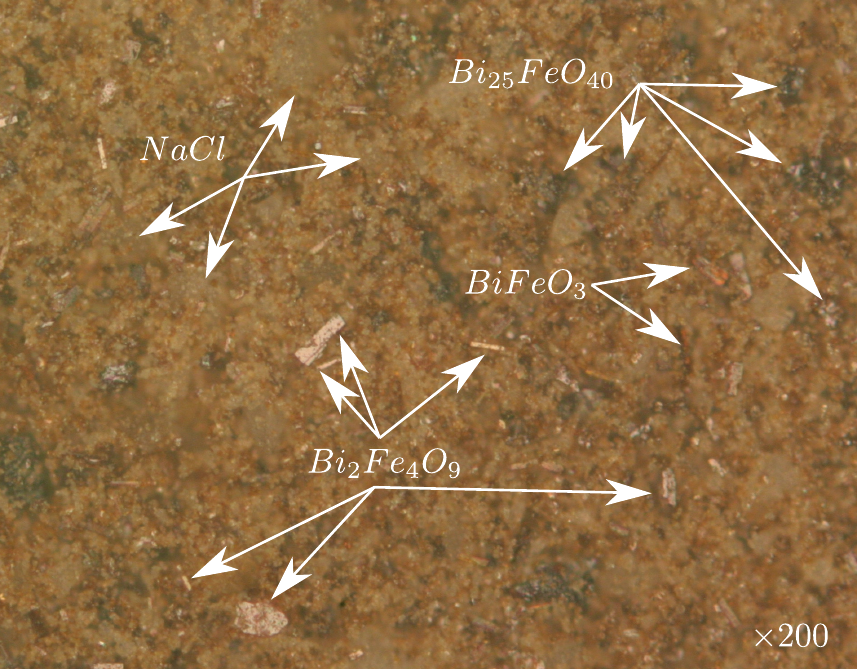}
\par\end{centering}
\caption{\label{fig:An-optical-microscope-6.8NaCl}An optical microscope image
of the synthesized and ground material with $6.5\,weight\%$ $NaCl$
prior to sintering. An example of the various phases that can be observed
are pointed to using arrows.}
\end{figure}

We now compare the crystallized compositions of the two synthesized
materials (with $Bi_{2}O_{3};\,NaCl$ flux and $Bi_{2}O_{3}$ flux)
after calcination and grinding. Figure \ref{fig:An-optical-microscope-6.8NaCl}
illustrates the ground material with $6.5\,weight\%$ $NaCl$.
X-ray diffraction measurements of the ground material with NaCl in
the flux show the following weight fraction of the crystalline phases
(Table \ref{tab:Weight-fraction-NaCl-flux})

\begin{table}[H]
\centering{}%
\begin{tabular}{llr@{\extracolsep{0pt}.}lr@{\extracolsep{0pt}.}lr@{\extracolsep{0pt}.}lr@{\extracolsep{0pt}.}lr@{\extracolsep{0pt}.}lr@{\extracolsep{0pt}.}lr@{\extracolsep{0pt}.}l}
\toprule 
Compound & \multicolumn{3}{l}{Weight fraction {[}\%{]}} & \multicolumn{2}{c}{$a\,[\mathring{A}]$} & \multicolumn{2}{c}{$b\,[\mathring{A}]$} & \multicolumn{2}{c}{$c\,[\mathring{A}]$} & \multicolumn{2}{c}{$\alpha\,[\degree]$} & \multicolumn{2}{c}{$\beta\,[\degree]$} & \multicolumn{2}{c}{$\gamma\,[\degree]$}\tabularnewline
\midrule
$Bi_{2}Fe_{4}O_{9}$ & ~~~~~~~~~~ & 27&4 & 7&96623  & 8&43444  & 5&99638  & 90&000 & 90&000 & 90&000\tabularnewline
$Bi_{22}Fe_{2}O_{36}$ &  & 26&3 & 10&13167 & 10&13167 & 10&13167 & 90&000 & 90&000 & 90&000\tabularnewline
$Bi_{25}FeO_{40}$ &  & 24&4 & 10&15868  & 10&15868  & 10&15868  & 90&000 & 90&000 & 90&000\tabularnewline
$BiFeO_{3}$ &  & 14&6 & 5&63427 & 5&63427 & 13&78349  & 90&000 & 90&000 & 120&000\tabularnewline
$\alpha-BiFeO_{3}$ &  & 7&3 & 5&62267 & 5&62936 & 5&62796 & 59&330 & 59&350 & 59&380\tabularnewline
\bottomrule
\end{tabular}\caption{\label{tab:Weight-fraction-NaCl-flux}Weight fraction of crystalline
phases in clacinated BFO $Bi_{2}O_{3};\,NaCl$ based flux}
\end{table}
As we can expect, we see that the prevailing phases are the mullite
(27.4\%) and sillenite (24.4\%). An additional, less commonly reported
compound $Bi_{22}Fe_{2}O_{36}$, was observed as well. This compound
was suggested by Mel'Nikova et al. \cite{mel2014development} who
developed a method of determining the composition of sillenites using
XRD data. Murakami \cite{murakami2018batio3} working on $BaTiO_{3}-BiFeO_{3}$
compounds, reports detecting the compound but does not dwell on the
subject. We assume that it is one of the metastable phases which is
attained due to the existence of the sodium ions, acting as spectator
ions. A smaller fraction (14.6\%) of rhombohedral $BiFeO_{3}$ phase
exists with an additional 7.3\% (2:1 ratio) of the triclinic (P1)
$\alpha-BiFeO_{3}$. The $\alpha-BiFeO_{3}$ results obtained, corresponds
with the lattice parameters reported by Wang et al. \cite{wang2013structure}.
It is interesting to notice that even though the NaCl did not react
with the material displaying obvious segregation it was found to be
in an amorphous state.

\begin{figure}[H]
\centering{}\includegraphics[scale=0.7]{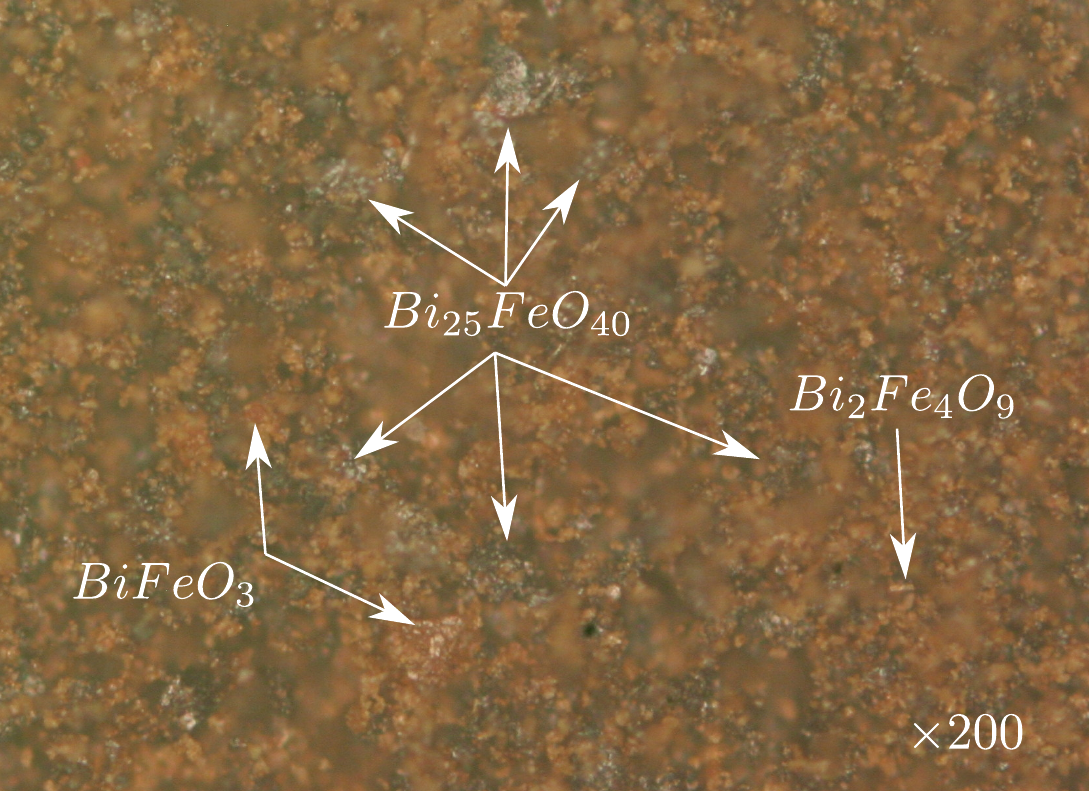}\caption{\label{fig:Optical-microscope-image-Bi2O3-flux}Optical microscope
image of the calcinated and ground $Bi_{2}O_{3}$ based flux}
\end{figure}

Figure \ref{fig:Optical-microscope-image-Bi2O3-flux} displays an
optical microscope image of the second calcinated and ground $Bi_{2}O_{3}$
based flux.
while the orthorhombic mullite structure is hardly observed in the
figure, one can easily notice the considerable increase in the metallic
gray areas signifying the sillenite compound. Table \ref{tab:Weight-fraction-of-Bi2O3-flux}
displays the weight fraction of the crystalline phases of the calcinated
flux materials that do not include the sodium chloride.
\begin{table}[H]
\begin{centering}
\begin{tabular}{llr@{\extracolsep{0pt}.}lr@{\extracolsep{0pt}.}lr@{\extracolsep{0pt}.}lr@{\extracolsep{0pt}.}lr@{\extracolsep{0pt}.}lr@{\extracolsep{0pt}.}lr@{\extracolsep{0pt}.}l}
\toprule 
Compound & \multicolumn{3}{l}{Weight fraction {[}\%{]}} & \multicolumn{2}{c}{$a\,[\mathring{A}]$} & \multicolumn{2}{c}{$b\,[\mathring{A}]$} & \multicolumn{2}{c}{$c\,[\mathring{A}]$} & \multicolumn{2}{c}{$\alpha\,[\degree]$} & \multicolumn{2}{c}{$\beta\,[\degree]$} & \multicolumn{2}{c}{$\gamma\,[\degree]$}\tabularnewline
\midrule
$Bi_{25}FeO_{40}$ &  & 68&9 & 10&15600  & 10&15600  & 10&15600  & 90&000 & 90&000 & 90&000\tabularnewline
$BiFeO_{3}$ &  & 25&7 & 5&57801  & 5&57801  & 13&86400  & 90&000 & 90&000 & 120&000\tabularnewline
$Bi_{2}Fe_{4}O_{9}$ & ~~~~~~~~~~ & 5&39 & 7&97971  & 8&43282  & 6&00463  & 90&000 & 90&000 & 90&000\tabularnewline
\bottomrule
\end{tabular}
\par\end{centering}
\caption{\label{tab:Weight-fraction-of-Bi2O3-flux}Weight fraction of crystalline
phases in calcinated BFO flux materials without sodium chloride}
\end{table}

Table \ref{tab:Weight-fraction-of-Bi2O3-flux} shows a similar percentage
of mullite to the material in which the parasitic phases were not
reduced (Lu \cite{lu2011phase}). In this molar ratio we do expect
the sillenite to be much more abundant as it is one of the two stable
phases. 

\subsection{Sintering}

\begin{figure}[H]
\centering{}\includegraphics[scale=0.7]{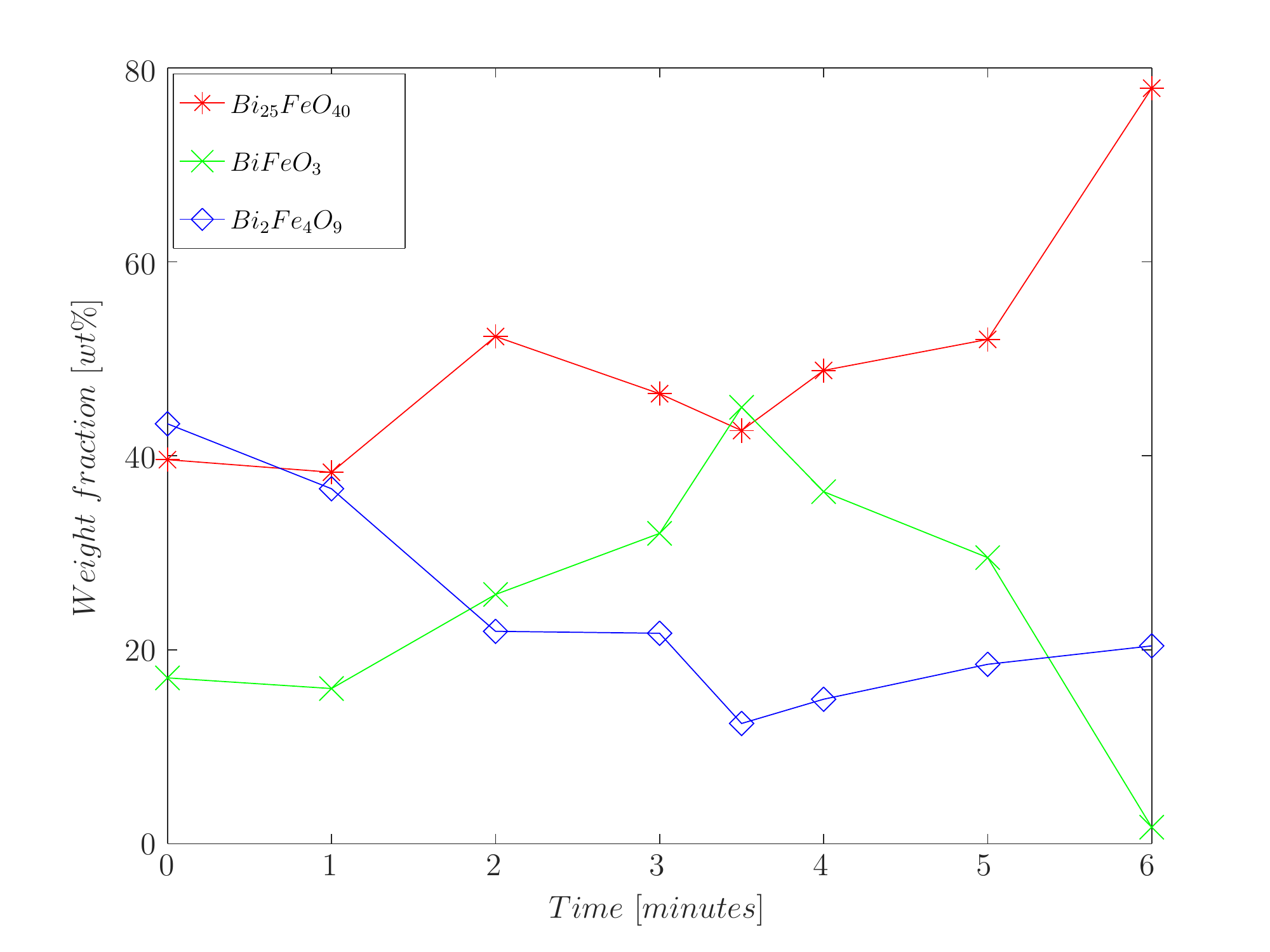}\caption{\label{fig:Weight-fraction-of-no-NaCl}Weight fraction of the different
compounds resulting from the solid state reaction of the quenched
disks at $880\degree C$ based on the reaction time.}
\end{figure}

Prior to sintering and in order test the stable phases, the materials
were placed for 16 hours at $600\degree C$ thus increasing the stable
phases at that temperature range, i.e. the mullite and sillenite phases.
The reaction trend gives us a better indication which are the stable
phases. As the various articles using rapid sintering \cite{mukherjee1971kinetics,wang2004room,awan2011synthesis}
reported placing the pressed disks within the furnace between 400
and 450 seconds, the samples were placed in the furnace up to 8 minutes
and then rapidly cooled by extracting them from the furnace and letting
them cool on a piece of ceramic at room temperature. Figure \ref{fig:Weight-fraction-of-no-NaCl}
illustrates the results obtained from the 56/44 mole\% $Bi_{2}O_{3}-Fe_{2}O_{3}$
solid state compounds based on XRD measurements.
At this specific $Bi_{2}O_{3}/Fe_{2}O_{3}$ ratio, the two expected
phases that are known to coexist are the BFO phase and the sillenite
phase. The results show that up to approximately 3.5 minutes the $BiFeO_{3}$
phase increased to about 50\% at the expense of both the $Bi_{25}FeO_{40}$
and $Bi_{2}Fe_{4}O_{9}$ phases, after which it experiences a fast
decline to undetectable levels. After 3.5 minutes while the BFO decomposes
the level of the mullite increases slightly and the sillenite becomes
the dominant stable phase. This does not correspond with the accepted
phase diagram and hints that some impurities from the crucible may
have altered the composition or that at $880\degree C$ $Bi_{25}FeO_{40}$
is the dominant phase. 

\begin{figure}[H]
\centering{}\includegraphics[scale=0.7]{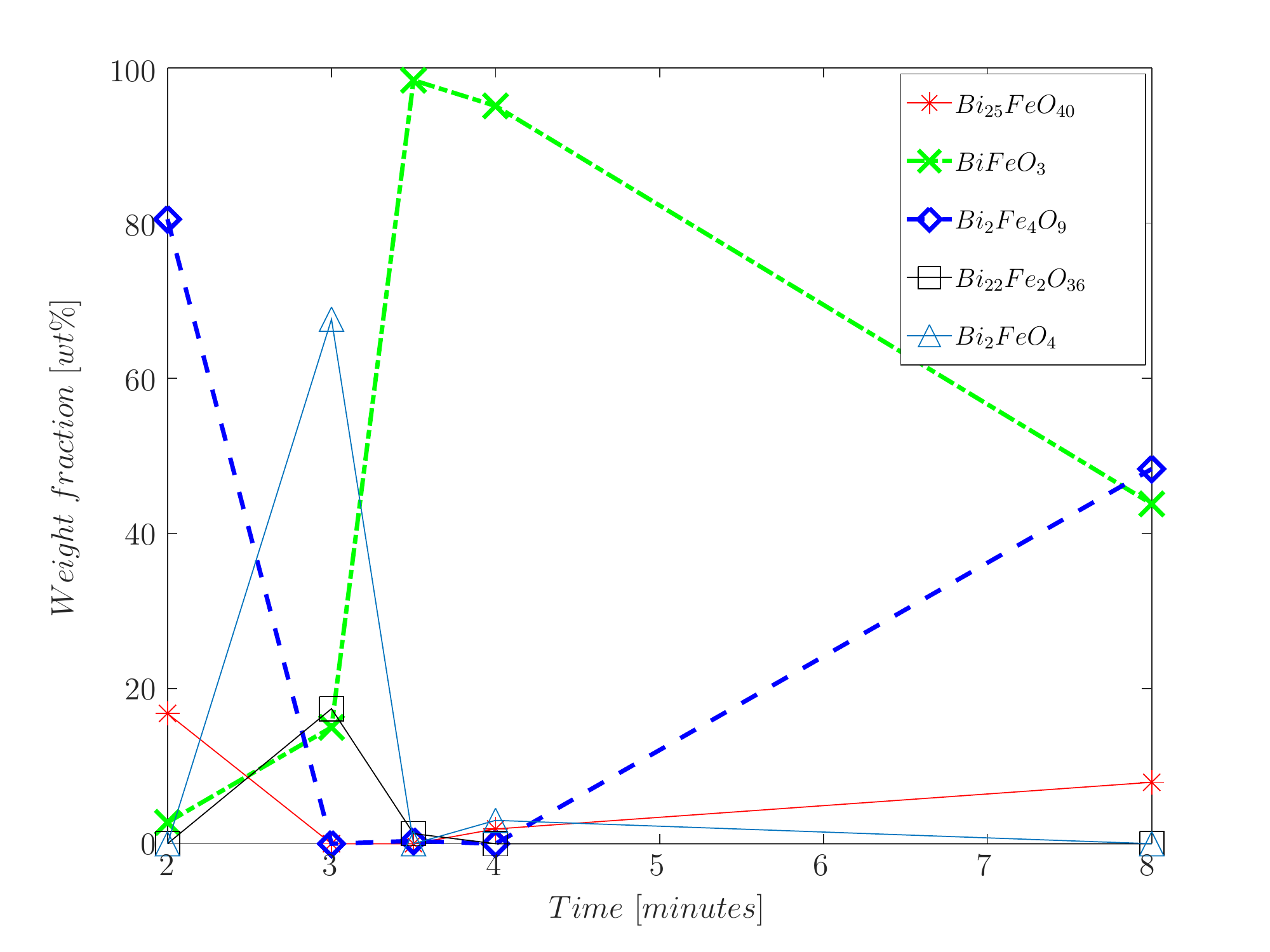}\caption{\label{fig:Weight-fraction-of-NaCl}Weight fraction of the different
compounds obtained by the $56Bi_{2}O_{3}\cdot44Fe_{2}O_{3}+6.5wt\%\,NaCl$
composition heated and quenched disks at $880\degree C$ based on
the reaction time.}
\end{figure}

We repeated the process described above, with the $Bi_{2}O_{3}-Fe_{2}O_{3}-NaCl$
composition. The weight fraction of the different compounds obtained
by the system as a function of time is described in figure \ref{fig:Weight-fraction-of-NaCl}.

Figure \ref{fig:Weight-fraction-of-NaCl} demonstrates the process
in which the NaCl (Melting point $801\degree C$) slowly, dissolves
the constituents and between 3-4 minutes suppressing the mullite and
sillenite phases while the $BiFeO_{3}$ temporarily dominates. At
this composition, the material is beyond the peritectic which for
longer periods will result in a liquid + $Bi_{2}Fe_{4}O_{9}$ similar
to the transition, for the 56/44 mole\% composition above $934\degree C$.

\section{Discussion}

Based on the accepted phase diagram \cite{Koizumi_1964,speranskaya1965phase,maitre2004experimental,palai2008beta,lu2011phase},
and Gibbs energy of formation \cite{selbach2009thermodynamic}, during
the synthesis of materials and solid state reactions above $767\degree C$
the $Bi_{2}Fe_{4}O_{9}$ phase, should not be a stable phase. This
is true as long as the $Bi_{2}O_{3}/Fe_{2}O_{3}$ molar ratio is above
1:2. Contrary to what has just been stated, experience has shown that
for long enough periods of time the mullite phase does form as a stable
phase. As a result, rapid heating and cooling techniques are a common
practice for obtaining a high weight fraction of BFO ceramics. Lu
et. al. have shown that for highly controlled compositions in gold
crucibles at a temperature of $850\degree C$, BFO is the prevalent
stable phase. Given impurities it will slowly decompose to other phases
primarily to mullite. At temperatures above $855\degree C$ even the
highly controlled compositions tend to decompose over a long period
of time. This has to be taken into account if one attempts to grow
bulk crystals even for highly controlled environments, one is limited
below $850\degree C$ for bulk BFO crystal growth. As the goal of
this work was sintering targets for sputtering machines, we were interested
in BFO phase stability and the adequate sintering time during the
sintering process. We compared a 56:44 mole\% $Bi_{2}O_{3}-Fe_{2}O_{3}$
composition to a $56Bi_{2}O_{3}\cdot44Fe_{2}O_{3}+6.5wt\%\,NaCl$
composition. We demonstrated that at $880\degree C$ after 3.5 minutes
in the composition which does not include NaCl only approximately
50\% of the material has transformed into BFO. At the same time using
the composition that includes NaCl as a flux, 98.5\% of the material
was in the BFO phase totally suppressing any other phase. The importance
of this finding is that it is relatively simple to remove the NaCl
from the sintered disk. This is done by rinsing the disk in water.
One drawback of this method is that it leaves the disk with pitting
and cavities.

\bibliographystyle{unsrt}
\bibliography{SinteredBFOTargetsPreprint}

\end{document}